\pdfoutput=1
\documentclass[11pt]{article}

\usepackage[final]{acl}

\usepackage{times}
\usepackage{latexsym}

\usepackage[T1]{fontenc}
\usepackage[utf8]{inputenc}

\usepackage{microtype}

\usepackage{inconsolata}

\usepackage{graphicx}

\usepackage{amsmath}
\usepackage{amsfonts}
\usepackage{hyperref}
\usepackage{multirow}
\usepackage{booktabs}
\usepackage{longtable}
\usepackage{caption}
\usepackage{subcaption}
\usepackage{graphicx}
\usepackage{soul}
\usepackage{float}

\definecolor{midnightgreen}{rgb}{0.0, 0.29, 0.33}

\title{Interpret and Control Dense Retrieval with Sparse Latent Features}

\author
{
    Hao Kang\\
    Carnegie Mellon University\\
    Pittsburgh, PA 15213\\
    \texttt{haok@andrew.cmu.edu}\\
\And
    Tevin Wang\\
    Carnegie Mellon University\\
    Pittsburgh, PA 15213\\
    \texttt{tevinw@andrew.cmu.edu}\\
\And
    Chenyan Xiong\\
    Carnegie Mellon University\\
    Pittsburgh, PA 15213\\
    \texttt{cx@cs.cmu.edu}
}

\begin{document}
    \maketitle
    \begin{figure*}[t]
    \centering
    \includegraphics[width=\linewidth]{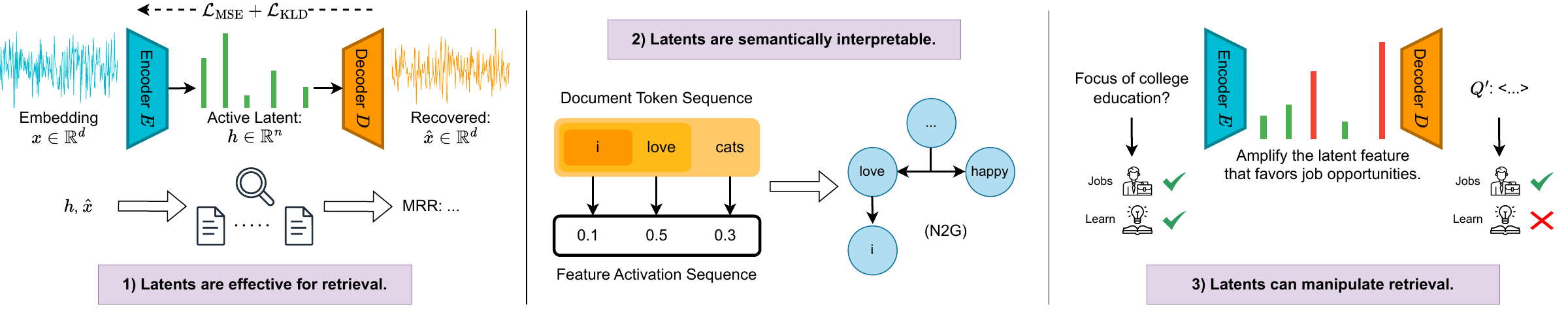}
    \caption{\label{figure:overview} An overview of our framework. We first train the $k$-sparse autoencoder with our retrieval-oriented contrastive loss, which produces sparse latent features that are effective for retrieval. Next, we interpret these latents using N2G approach and demonstrate controllability via retrieval on the manipulated embeddings.}
\end{figure*}

\begin{abstract}
Dense embeddings deliver strong retrieval performance but often lack interpretability and controllability. This paper introduces a novel approach using sparse autoencoders (SAE) to interpret and control dense embeddings via the learned latent sparse features. Our key contribution is the development of a retrieval-oriented contrastive loss, which ensures the sparse latent features remain effective for retrieval tasks and thus meaningful to interpret. Experimental results demonstrate that both the learned latent sparse features and their reconstructed embeddings retain nearly the same retrieval accuracy as the original dense vectors, affirming their faithfulness. Our further examination of the sparse latent space reveals interesting features underlying the dense embeddings and we can control the retrieval behaviors via manipulating the latent sparse features, for example, prioritizing documents from specific perspectives in the retrieval results.
\end{abstract}

\section{Introduction}

In the realm of information retrieval, dense embeddings derived from large language models (LLMs) have achieved state-of-the-art performances \cite{khattab2020colbert, reimers2019sentence}. While these representations offer remarkable accuracy in matching queries to documents, their ``black-box'' nature poses challenges in applications that demand transparency and control, such as retrieval in bias-sensitive tasks, where users may need to understand the rationale behind the retrieved results and adjust the process to ensure fairness.

In contrast, in bag-of-word base sparse retrieval, each dimension is a meaningful word, allowing users to see why certain documents are retrieved, and making it intuitive for users to revise their query keywords to control the retrieval results. Interpretability and controllability are important for building trust with users and facilitate the wide adoption of search technologies~\cite{croft2010search}.

In this paper, we present a novel approach that leverages sparse autoencoders (SAE) to interpret and control dense retrieval systems. Sparse autoencoders have recently been used to improve the interpretability of LLMs by transforming neuron activation patterns into sparse dictionaries \citep{bricken2023monosemanticity, templeton2024scaling}. We upgrade this approach to dense embeddings, incorporating a retrieval-oriented recovery loss which ensures the extracted sparse features remain faithful for retrieval, forming the basis of our interpretability analysis.

Our experiments demonstrate the success of this approach. Retrieval using the learned latent sparse features and their reconstructed embeddings both recover the majority of the original dense retrieval accuracy on the \textsc{MsMarco} and \textsc{Beir} benchmarks, ensuring that these features offer genuine interpretability rather than an illusion. Then we explore the interpretability of these sparse features with Neuron to Graph (N2G) approach \citep{foote2023neuron}, and discover that various fine-grained concepts have been captured in the latent sparse space.

To understand controllability through latent features, we conduct quantitative studies by amplifying query-relevant features, which successfully improved retrieval accuracy on the manipulated embeddings, both on the query side and the document side. Then, we perform case studies on multi-perspective queries and confirm that selectively manipulating sparse features from a specific perspective causes the reconstructed embeddings to prioritize documents from that perspective during retrieval. Our source code and extracted features are available at GitHub \footnote{\href{https://github.com/cxcscmu/embedding-scope}{https://github.com/cxcscmu/embedding-scope}}.

    \section{Methodology} \label{section:methodology}

In this section, we describe the methodology used to train the sparse autoencoder with our retrieval-oriented recovery loss.

As illustrated in Figure \ref{figure:overview}, for an embedding vector $x\in\mathbb{R}^d$, we employ the $k$-sparse autoencoder as proposed in \citet{makhzani2013k}, which  controls the number of active latent features using the TopK activation function. The encoder and decoder are described in Equation~\ref{eq:encoder_and_decoder}, where $n$ denotes the latent dimension for $W_\text{enc}\in\mathbb{R}^{n\times d}$. The reconstructed embedding is represented by $\hat{x}\in\mathbb{R}^d$.

\begin{equation}
\label{eq:encoder_and_decoder}
\begin{aligned}
    h&=\text{TopK}(W_\text{enc}(x-b_\text{dec})+b_\text{enc})\\
    \hat{x}&=W_\text{dec}h+b_\text{dec}\\
\end{aligned}
\end{equation}

Building on previous efforts to extract interpretable features from LLMs \citep{gao2024scaling, bricken2023monosemanticity, lieberum2024gemma}, we incorporate mean-squared error (MSE) as part of the training objective for reconstruction. By minimizing the squared differences, MSE forces each dimension of the reconstructed embedding to closely approximates the original value.

However, the focus of MSE is to minimize the error for individual points in the embedding space. It does not explicitly account for the relative positioning. For information retrieval, embeddings are typically divided into queries and documents, with the need to effectively capture the relevance between a query and its associated documents.

Therefore, we employ contrastive learning via Kullback–Leibler divergence (KLD) to ensure that the distribution of reconstructed query and document embedding aligns with the original \cite{xiong2020answering, liu-etal-2022-dimension}. The formulation of the loss function is presented in Equation~\ref{eq:loss_function}, where $q$ represents the query embedding, $D^+$ denotes the relevant documents, and $f(q, d)$ computes the retrieval score, such as dot product.

\begin{equation}
\label{eq:loss_function}
\begin{aligned}
    \mathcal{L}_\text{KLD}=\sum_q&\sum_{d\in D^+}P(q,d)\times\log{\frac{P(q,d)}{P(\hat{q},\hat{d})}}\\
    \text{where}&~P(q,d)=\frac{e^{f(q,d)}}{\sum_{D^+}e^{f(q,d)}}\\
\end{aligned}
\end{equation}

In short, the $k$-sparse autoencoder is trained with MSE for accurate reconstruction and KLD to preserve the query-document relationship.

    \begin{table*}[t]
\small
\centering
\caption{\label{table:evaluation} Reconstruction evaluation of sparse latent features and the reconstructed embeddings learned by our $k$-sparse autoencoder from the \textsc{Bge} model. MSE measures the embedding differences between original and reconstructed embeddings. Results for the alternative \textsc{MiniCPM} embedding model can be found in Appendix \ref{appendix:role-of-base-embedding}.}
\begin{tabular}{l|cccc|cccc}
    \toprule
    &\multicolumn{4}{c|}{\textsc{MsMarco}}&\multicolumn{4}{c}{\textsc{Beir}}\\
    &\textbf{MSE}&\textbf{MRR}&\textbf{P@10}&\textbf{R@10}&\textbf{MSE}&\textbf{MRR}&\textbf{P@10}&\textbf{R@10}\\
    \hline
    Original&--&0.3605&0.0649&0.6211&--&0.3699&0.0891&0.5415\\
    \hline
    Sparse Latent (K=32)&--&0.2721&0.0507&0.4869&--&0.2420&0.0581&0.3590\\
    Sparse Latent (K=64)&--&0.3062&0.0564&0.5406&--&0.2923&0.0708&0.4212\\
    Sparse Latent (K=128)&--&\textbf{0.3306}&\textbf{0.0601}&\textbf{0.5760}&--&\textbf{0.2981}&\textbf{0.0735}&\textbf{0.4461}\\
    Reconstructed (K=32)&0.00022&0.2984&0.0552&0.5291&0.00043&0.2549&0.0619&0.3768\\
    Reconstructed (K=64)&0.00017&0.3194&0.0583&0.5589&0.00033&0.2913&0.0721&0.4361\\
    Reconstructed (K=128)&\textbf{0.00011}&\textbf{0.3455}&\textbf{0.0626}&\textbf{0.5991}&\textbf{0.00019}&\textbf{0.3407}&\textbf{0.0818}&\textbf{0.4954}\\
    \bottomrule
\end{tabular}
\end{table*}

\section{Experiments} \label{section:experiment}

This section outlines the training procedures for the $k$-sparse autoencoder and our experiments on interpretability and controllability.

\textbf{Training Procedures.} We train the autoencoder on top of the base-sized \textsc{Bge} model \footnote{\href{https://huggingface.co/BAAI/bge-base-en-v1.5}{https://huggingface.co/BAAI/bge-base-en-v1.5}}, which was trained on diverse tasks such as retrieval, classification, and semantic similarity \cite{xiao2023c}. Embeddings are generated from the \textsc{MsMarco} dataset, containing 8.8M passages for retrieval tasks \citep{bajaj2016ms}. Details of the training hyperparameters are available in Appendix~\ref{appendix:training}.

For evaluation, we first calculate MSE on the validation queries and their relevant documents. We then perform dense retrieval on the reconstructed embeddings and sparse dot product retrieval on the latent features. Reported metrics include mean reciprocal rank (MRR), precision at rank 10 (P@10), and recall at rank 10 (R@10).

For generalizability on diverse retrieval tasks, we additionally evalute the sparse autoencoder on datasets from the \textsc{Beir} benchmark, such as \textsc{TrecCovid}, \textsc{NaturalQuestions}, and \textsc{DBPediaEntity} \citep{kwiatkowski2019natural, hasibi2017dbpedia, thakur2021beir}. Additionally, we investigate the impact of the base embedding by applying our approach to an alternative embedding model, \textsc{MiniCPM} \footnote{\href{https://huggingface.co/openbmb/MiniCPM-Embedding}{https://huggingface.co/openbmb/MiniCPM-Embedding}} \citep{hu2024minicpm}.

\textbf{Interpretability Study.} To assess interpretability, we generate N2G explanations \citep{foote2023neuron}. N2G provides an automated approach to interpret the behavior of individual neurons by converting their activations into graph-based representations. It identifies the most relevant tokens that strongly activate a neuron and focuses on them by pruning the surrounding, less relevant context. This process isolates the essential patterns that contribute to the neuron's activation. 

Additionally, N2G enriches the dataset by replacing key tokens with high-probability substitutes, generating variations that maintain high activation levels. By doing so, the method captures a broader and more nuanced understanding of the neuron's behavior, revealing how it responds to different inputs while maintaining its core functionality. This combination of pruning and augmentation ensures that the interpretability of each neuron is both concise and comprehensive \cite{foote2023neuron}. 

For each feature, we create a training set of 512 samples by selecting the highest-activating documents. We then perform forward passes on prefix sequences to extract activation sequences, which are input into N2G to construct trie representations for each feature. \textsc{Gpt-4o-mini} is used to interpret each trie's semantic meaning.

\textbf{Controllability Study.} In the controllability experiments, we explore how amplifying sparse latent features based on relevance can influence retrieval. The experiments involve manipulating document and query embeddings.

For document manipulation, we amplify the latent feature of relevant documents in the dimension corresponding to the highest activation of each query. The modified latent features are then decoded to reconstruct the document embeddings for retrieval. For query manipulation, we amplify query features in the dimension most activated by relevant documents. A grid search determines the appropriate amplification level, starting with the smallest value of latent features at 0.0004, incremented by a factor of 2 each step.

On the other hand, we explore binary perspective queries, structured to have two distinct categories of potential document matches in our control experiments. By amplifying the latent features associated with these categories, we assess whether manipulating a particular feature leads to a greater prevalence of one category over the other during retrieval on the reconstructed embeddings.

    \section{Evaluation}

In this section, we present the evaluated results for each experiment in Section~\ref{section:experiment} and discuss the underlying insights that are critical for our findings.

\subsection{Retrieval Performance}

The final results in Table~\ref{table:evaluation} confirm the robustness of the reconstruction. With K=128 active features in the latent space, the MSE on the \textsc{MsMarco} dataset is 0.0001, and the MRR reaches 0.3455, closely aligning with the original score of 0.3605. Notably, the features extracted by the sparse autoencoder also prove valuable for retrieval, achieving an MRR of 0.3306. This utility strengthens our confidence that the interpretability analysis provides genuine insights rather than illusory interpretations.

\begin{figure}[t]
\centering
\begin{subfigure}[b]{0.49\linewidth}
    \centering
    \includegraphics[width=\linewidth]{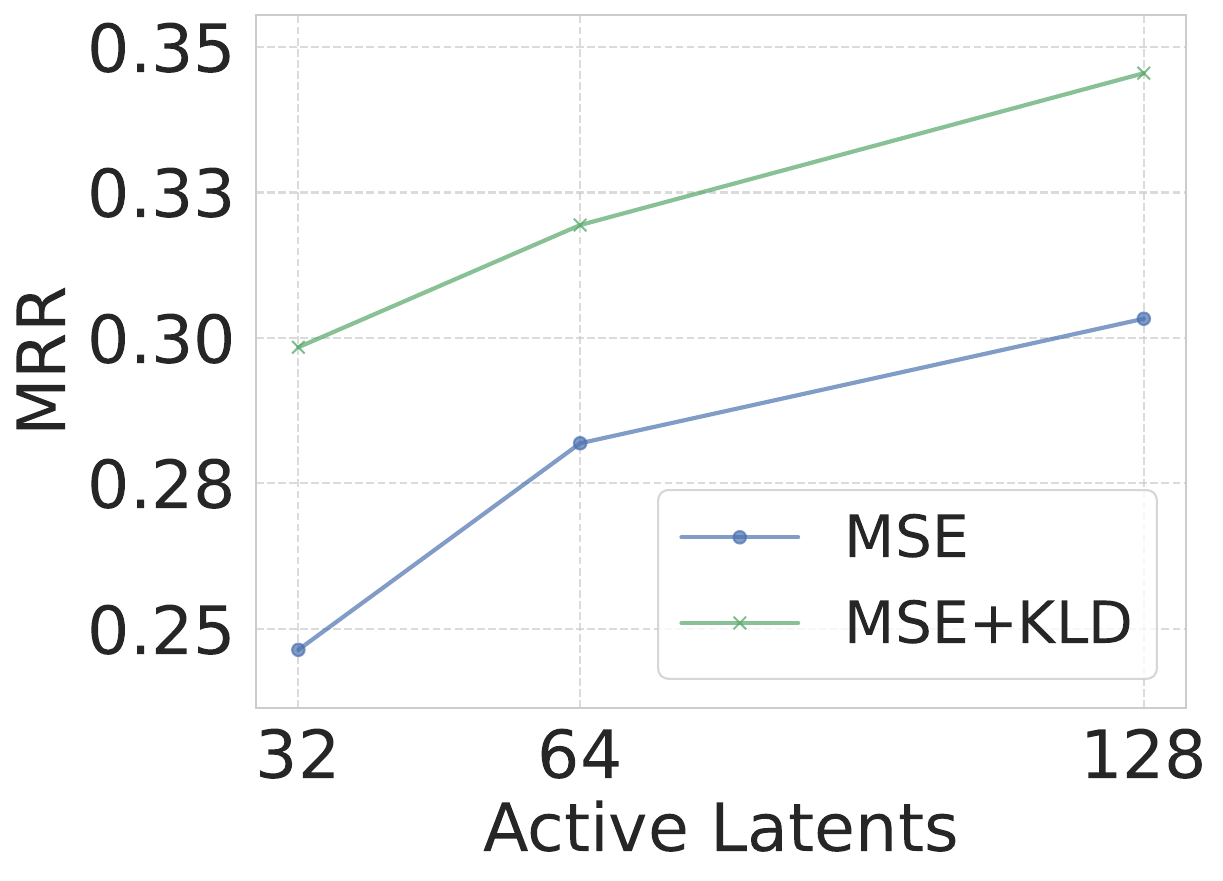}
    \caption{\textsc{MsMarco} (Rec.)}
    \label{fig:ablation1}
\end{subfigure}
\hfill
\begin{subfigure}[b]{0.49\linewidth}
    \centering
    \includegraphics[width=\linewidth]{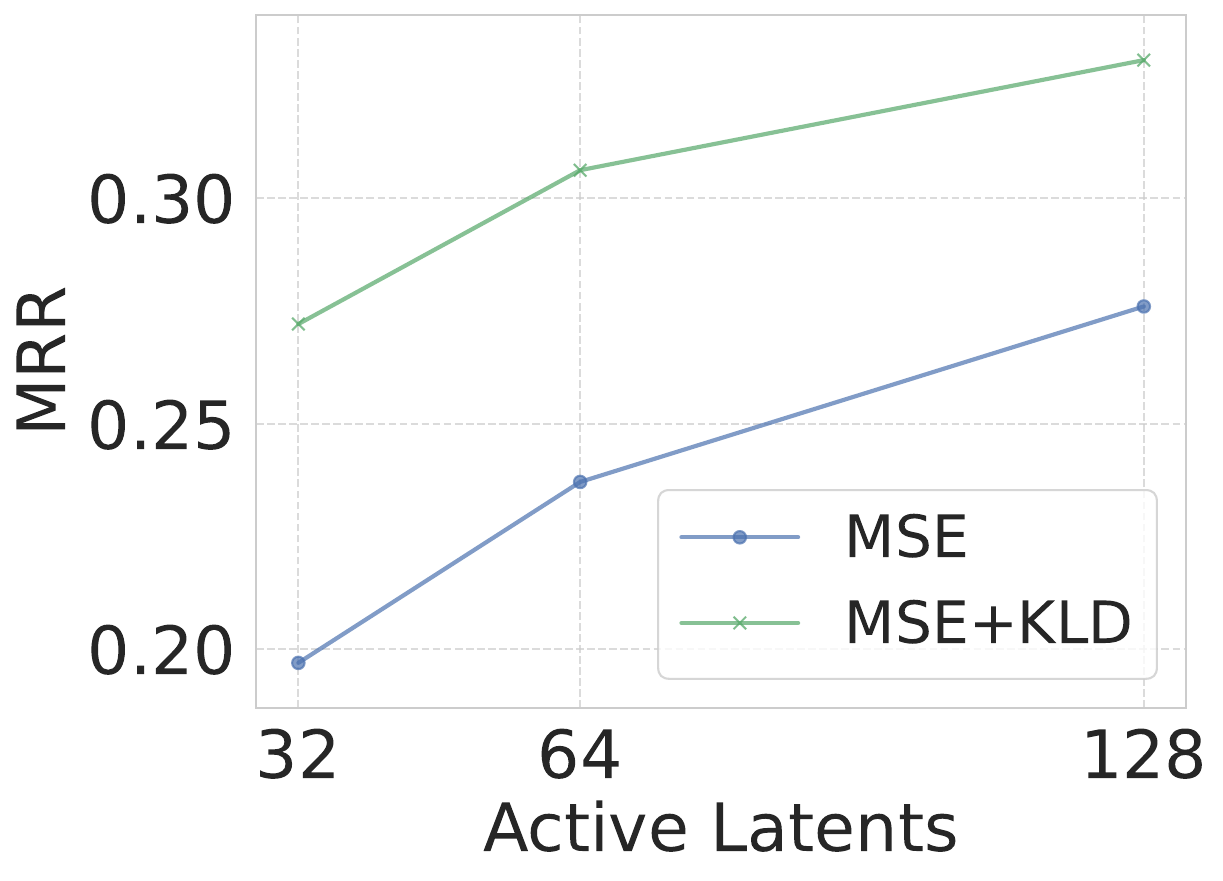}
    \caption{\textsc{MsMarco} (Spr.)}
    \label{fig:ablation2}
\end{subfigure}
\caption{\label{figure:ablation} Retrieval performance of reconstructed (Rec.) embeddings and the sparse latent features (Spr.) before and after the contrastive loss KLD is applied on \textsc{MsMarco} using \textsc{Bge} as the embedding model. Results on \textsc{Beir} can be found in Appendix \ref{appendix:ablation-study}.}
\end{figure}

We further assessed the impact of contrastive loss through an ablation study, comparing models trained with MSE alone against those incorporating contrastive loss. All other conditions were kept identical to ensure a fair comparison. As presented in Figure~\ref{figure:ablation}, the model trained with contrastive loss consistently outperforms the baseline across all latent dimensions. Notably, retrieval on sparse features improves the MRR to 0.3306, compared to 0.2760. Even though both models experience performance drop for retrieval on the \textsc{Beir} dataset, models trained with contrastive loss demonstrate better resilience, suggesting stronger robustness across diverse retrieval tasks.

\subsection{Interpretability Study}

\begin{figure}[t]
    \centering
    \small
    \begin{subfigure}{0.49\linewidth}
        \centering
        \includegraphics[width=\linewidth]{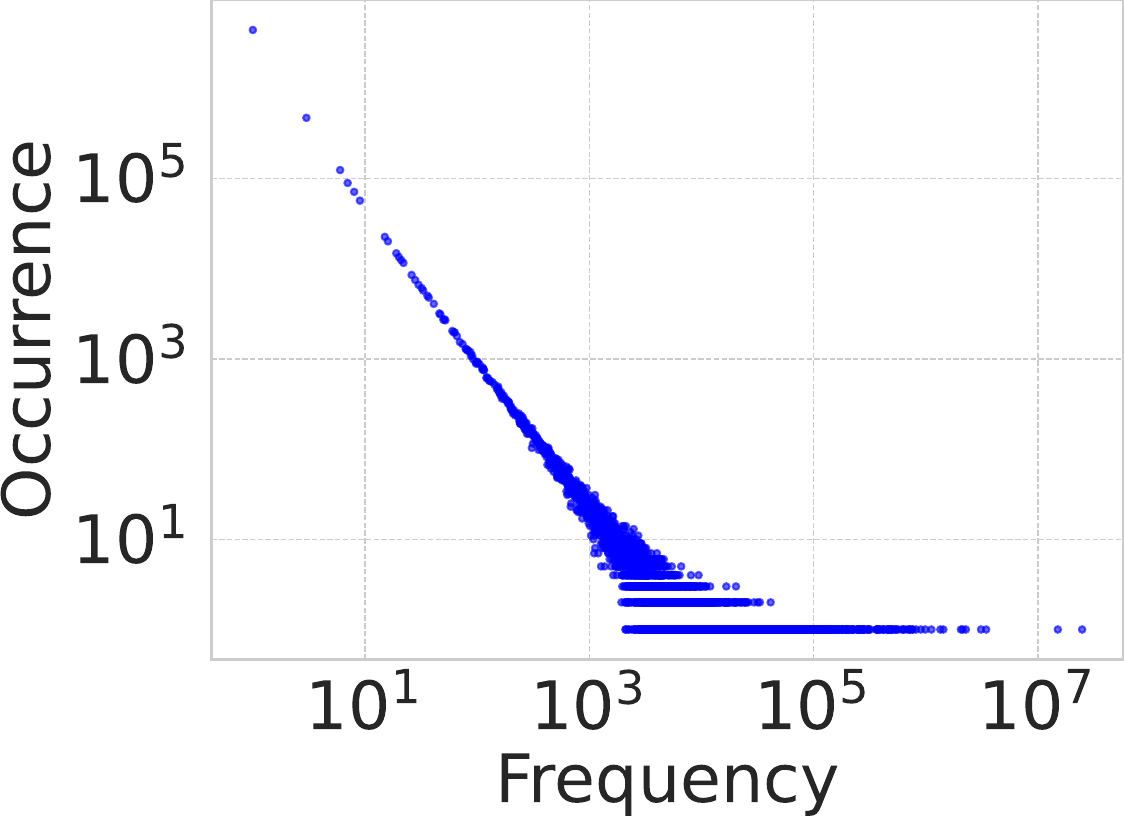}
        \caption{Unigram Bag-of-Words}
        \label{figure:unigram_bag_of_words}
    \end{subfigure}
    \hfill
    \begin{subfigure}{0.49\linewidth}
        \centering
        \includegraphics[width=\linewidth]{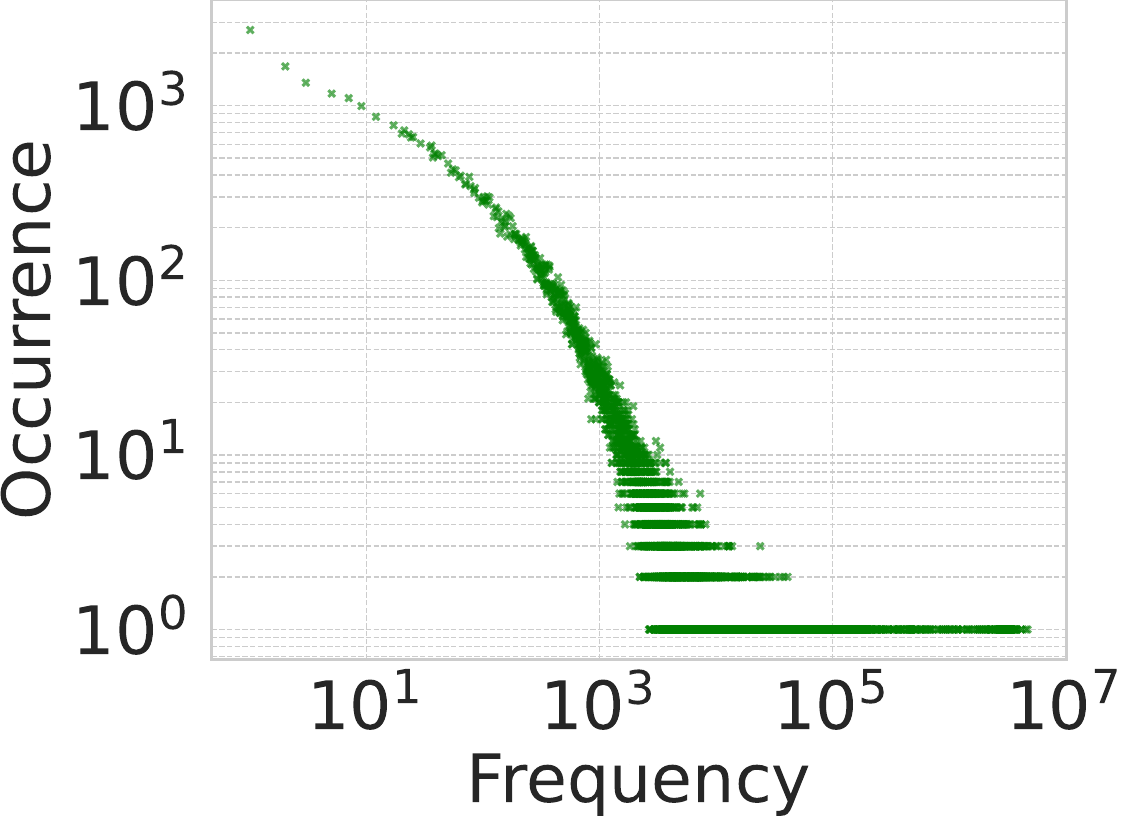}
        \caption{Sparse Latent Features}
        \label{figure:sparse_latent_features}
    \end{subfigure}
    \caption{\label{figure:frequency} Frequency distribution comparison between bag-of-words and sparse latent features in \textsc{MsMarco} using \textsc{Bge} as the embedding model. The high-frequency region is characterized by a small number of words that occur with extreme regularity, whereas the low-frequency region consists of a large proportion of words that appear only a limited number of times throughout the dataset.}
\end{figure}

\begin{table}[t]
\small
\centering
\caption{\label{table:regional_features} Examples of sparse latent features using \textsc{Bge} as the embedding model explained by N2G from different parts of the frequency distribution.}
\begin{tabular}{r|l}
    \toprule
    \textbf{Region}&\textbf{Description from N2G}\\
    \midrule
    \multirow{3}{*}{Head}
    &media, production, television, entertainment\\ \cline{2-2}
    &fashion, appearance, behavior, transformation\\ \cline{2-2}
    &opera, drama, music, performance, composer\\
    \midrule
    \multirow{3}{*}{Torso}
    &korea, seoul, music, culture, tourism\\  \cline{2-2}
    &sports, injuries, protocols, regulations\\ \cline{2-2}
    &location, community, development, services\\
    \midrule
    \multirow{3}{*}{Tail}
    &health, pain, injury, trauma, disorders\\ \cline{2-2}
    &growth, improvement, learning, strategy\\ \cline{2-2}
    &finance, investment, market, companies\\
    \bottomrule
\end{tabular}
\end{table}

\begin{table}[t]
\small
\centering
\caption{Top activated features using \textsc{Bge} as the embedding model from the document ``A few people reported that they paid their attorney as little as \$50 per hour, and a few reported paying as much as \$400 to \$650 per hour. But the vast majority paid between \$150 and \$350 per hour, with \$250 being the most commonly reported fee. The survey asked respondents about a number of things, including: 1 how much their divorce attorney charged per hour. 2 how much their divorce cost. 3 the number of issues that they resolved out of court and in court. 4 whether their spouse contested the case. 5 how long the divorce took from start to finish.''}
\begin{tabular}{l@{\hskip 5pt}l}
    \toprule
    \textbf{Description from N2G} \\
    \midrule
    1. cost, pricing, expenses, rates, income \\
    2. time, duration, sleep, hours, minutes \\
    3. government, law, agencies, constitution, enforcement \\
    4. tennis, courts, wimbledon, justices, decisions \\
    5. health, anxiety, symptoms, stress, concerns \\
    \bottomrule
\end{tabular}
\label{table:document-features}
\end{table}

As illustrated in Figure~\ref{figure:frequency}, the learned sparse latent features also follow Zipf's law, but its  distribution is less head-heavy. This is interesting as top-ranking features in the bag-of-words model are often common stop words, but the sparse latent features may skip these stop words and capture fine-grained and conceptually meaningful categories. Representative feature examples extracted by N2G from different segments of the distribution are provided in Table~\ref{table:regional_features}, while the top activated features for a sampled document in \textsc{MsMarco} dataset are detailed in Table~\ref{table:document-features}. Additional examples can be found in Appendix~\ref{appendix:interpretability}.

\subsection{Controllability Study}

\begin{figure}[t]
    \centering
    \small
    \begin{subfigure}{0.48\linewidth}
        \centering
        \includegraphics[width=\linewidth]{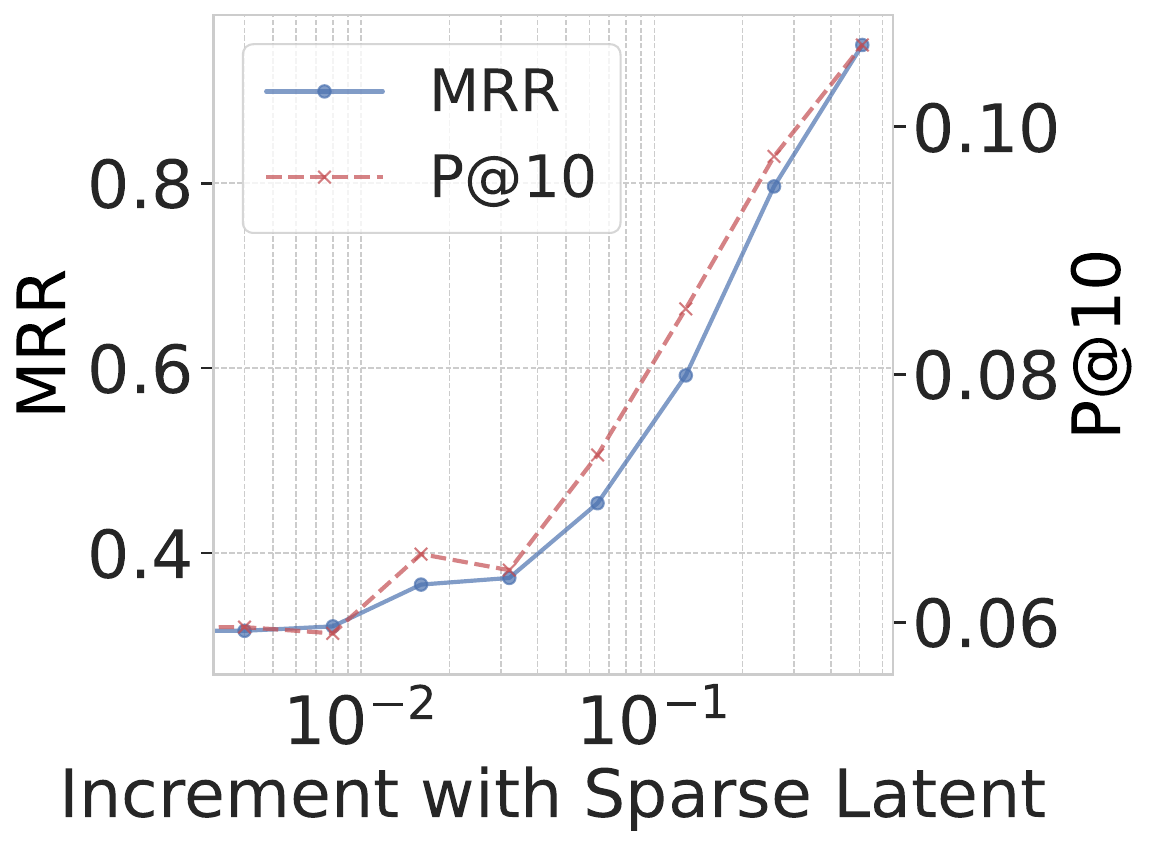}
        \caption{Manipulated Document}
        \label{fig:manipulated_document_retrieval}
    \end{subfigure}
    \hfill
    \begin{subfigure}{0.48\linewidth}
        \centering
        \includegraphics[width=\linewidth]{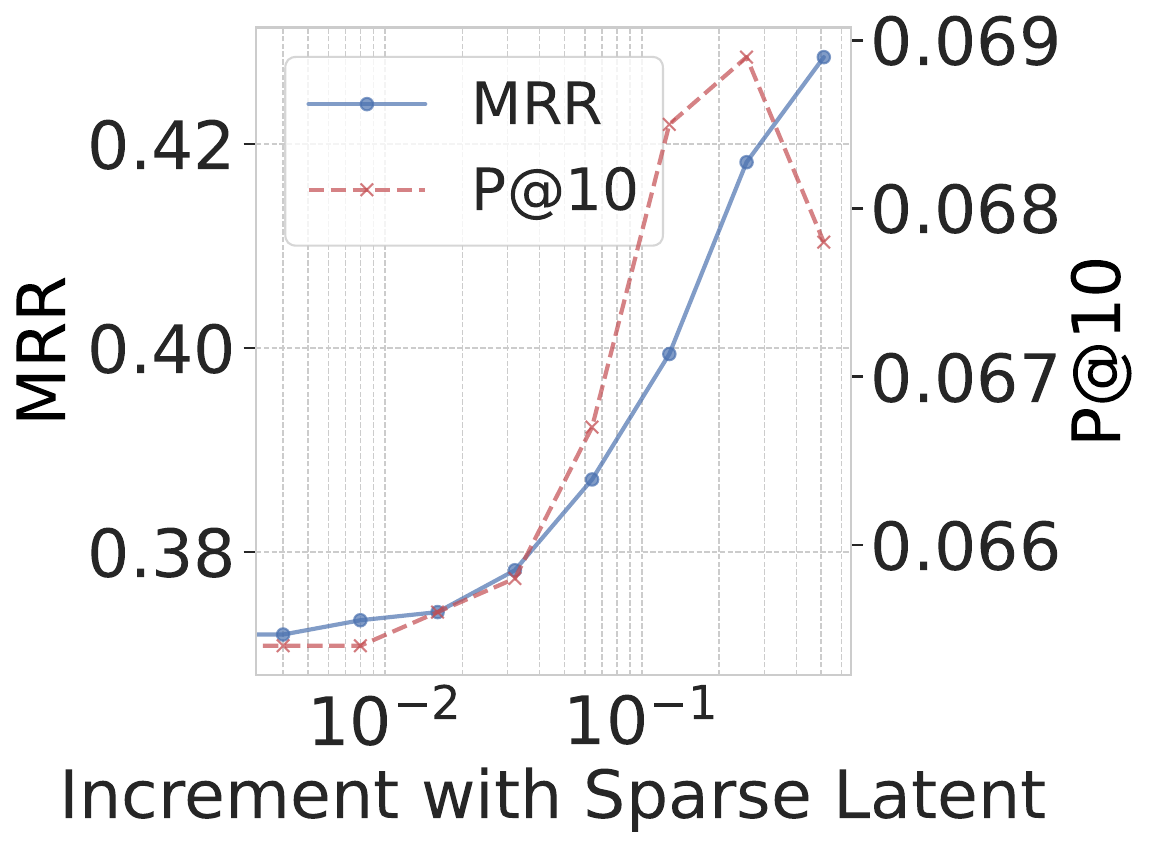}
        \caption{Manipulated Query}
        \label{fig:manipulated_query_retrieval}
    \end{subfigure}
    \caption{\label{figure:improvement} Improvement in retrieval scores on manipulated documents and queries by amplifying relevant sparse latent features across varying amounts using \textsc{Bge} as the embedding model. The x-axis is in logarithmic scale for better visualizing the trends since each step gets incremented by a factor of 2.}
\end{figure}

As shown in Figure~\ref{figure:improvement}, we observe a clear trend of improvement in both MRR and P@10 as the amplification of relevant sparse latent features increases. This demonstrates the controllability of latent features in influencing the retrieval process within the reconstructed embeddings. Specifically, as more relevance information is injected into the latent space, the retrieval scores improve. Notably, with document manipulation, the MRR reaches a peak value of 1.0 at the largest amplification level. It is also not surprising to see the performance drop on the query side when the manipulation is too strong---doubles the typically latent feature values---as it may break the reconstructed embedding.

\begin{table*}[t]
\small
\centering
\caption{\label{table:perspective:1} Features for the binary perspective query ``What is the primary focus of a university education?'' and the top result after dense retrieval on the reconstructed embeddings using \textsc{Bge} as the embedding model. Feature activations were amplified by 0.5. B/A displays the number of documents related to the feature before and after the amplification on $k=5$ retrieval.}
\begin{tabular}{c|p{37mm}|p{86mm}|c}
    \toprule
    \textbf{Feature ID} &\textbf{Description from N2G} & \textbf{Retrieved Document} & \textbf{B/A}\\ 
    \midrule
    84340  & employment, salary, wages, jobs, bonuses & ``...prepare people to work in various sectors of the economy or areas of culture...'' & 2/3\\
    \midrule
    179723 & growth, improvement, learning, strategy, development & ``...for students to own knowledge, hone capacities, develop personal and social responsibility...'' & 3/5\\
    \bottomrule
\end{tabular}
\end{table*}

Table~\ref{table:perspective:1} presents one example of controlling the retrieval results by manipulating the reconstructed query embeddings via the latent space. It shows that amplifying the targeted feature dimension effectively biases the retrieval results towards the corresponding perspective, i.e., ``job'' (84340) or ``learning'' (179723). This indicates that the learned faithful latent space provides a new mechanism to control the retrieval behavior which leads to many potential applications, for example, in enhancing the safety with human intervention in dense retrieval systems. Additional examples can be found in Appendix~\ref{appendix:controllability}.

    \section{Conclusion}

In this paper, we presented a novel method that applies sparse autoencoder to enhance the interpretability and controllability of dense embedding spaces in information retrieval. Our approach, which utilizes a retrieval-oriented contrastive loss function, ensures that the sparse features extracted remain faithful for interpretation. The experimental results demonstrate that our reconstructed embeddings maintain competitive retrieval accuracy, with sparse latent features proving to be both interpretable and controllably influential on retrieval outcomes. By enabling explicit manipulation of these sparse features, we provide a means to directly influence retrieval behaviors, offering a significant advantage for applications requiring transparent and adjustable retrieval mechanisms.

    \section{Limitations}

One limitation of this work is the potential for scaling. While the method demonstrates effectiveness, its scalability to larger embedding space remains to be explored. Additionally, although the sparse latent features offer strong evidence of interpretability and controllability, the relationship between these features and retrieval outcomes is still correlational, rather than causal. Thus, there is no guarantee that manipulating these features will always lead to the desired retrieval behavior. Lastly, while the sparse latent space approximates the performance of dense embeddings, it has not fully recovered the original retrieval performance, indicating room for further improvement.

    \bibliography{sections/References}
    \newpage
\appendix

\begin{table*}[t]
\small
\centering
\caption{
\label{table:perspective:2} 
Manipulation over for the binary perspective queries ``What is a key factor in the spread of infectious diseases?'' and ``What is a major influence on automotive emissions?'' by amplifying the perspective latent features using \textsc{BGE} as the embedding model.
}
\vspace{-0.2cm}
\begin{tabular}{c|p{32mm}|p{80mm}|c}
\toprule
    \textbf{Feature ID} &\textbf{Description from N2G} & \textbf{Retrieved Document} & \textbf{B/A} \\ 
    \midrule
    15678 & health, nutrition, immune, disease, metabolism & ``...1  Route of entry of the pathogen and the access to host regions that it gains. 2  Intrinsic virulence of the particular organism...''& 2/3\\
    53246 & demographics, migration, populations, countries, socioeconomic & ``...Learn how our modern way of life contributes to the spread and emergence of disease. 1  Globalization. 2  Climate Change. 3  Ecosystem Disturbances. 4 Poverty, Migration \& War...'' & 1/4\\
    \midrule
    142071 & climate, weather, precipitation, seasons, diversity   & ``... Major smog occurrences often are linked to heavy motor vehicle traffic, high temperatures, sunshine, and calm winds....'' & 2/5\\
    155875 & automotive, engineering, mechanics, combustion, manufacturing & ``...1  Driving and atmospheric conditions. 2  Mileage. 3  Vehicle age.  Type of spark plug electrode 1  material. Poor vehicle maintenance.  Poor quality 1  fuel. Damaged or worn sensors.  Dry-rotted or cracked vacuum hoses....''& 3/5\\ 
    \bottomrule
\end{tabular}
\end{table*}

\begin{table*}[t]
\small
\centering
\caption{
    \label{table:frequency-samples}
    Sparse latent features from the frequency distribution using \textsc{BGE} as the embedding model.
}
\vspace{-0.2cm}
\begin{tabular}{c|rl}
    \toprule
    \textbf{Region} & \textbf{Feature ID} & \textbf{Description from N2G}\\
    \midrule
    \multirow{6}{*}{Head}
    & 3 & media, production, television, entertainment\\
    & 24 & fashion, appearance, behavior, transformation\\
    & 30 & opera, drama, music, performance, composer\\
    & 58 & health, dignity, history, identity, inquiry\\
    & 82 & festival, country, music, education, rural\\
    & 86 & identity, culture, lifestyle, expression, community\\
    \midrule
    \multirow{6}{*}{Torso}
    & 28840 & korea, seoul, music, culture, tourism\\
    & 53784 & sports, injuries, protocols, regulations\\
    & 73817 & location, community, development, services\\
    & 91052 & meaning, significance, language, culture\\
    & 99785 & age, death, health, statistics, history\\
    & 194488 & weather, precipitation, climate, population\\ 
    \midrule
    \multirow{6}{*}{Tail}
    & 136995 & health, pain, injury, trauma, disorders\\
    & 179723 & growth, improvement, learning, strategy\\
    & 182171 & finance, investment, market, companies\\
    & 137124 & healthcare, assessment, professionals\\
    & 143764 & health, anatomy, surgery, body, women\\
    & 189083 & temperature, climate, weather, humidity\\
    \bottomrule
\end{tabular}
\end{table*}

\begin{table*}[t]
\small
\centering
\caption{
\label{table:query-features}
Top activated features from a subset of queries in \textsc{MsMarco} dataset using \textsc{BGE} as the embedding model.
}
\vspace{-0.2cm}
\begin{tabular}{c|rl}
    \toprule
    \textbf{Query Text} & \textbf{Feature ID} & \textbf{Description from N2G }\\
    \midrule
    \multirow{5}{*}{``what is prism in eyeglasses''}
    & 3125 & pattern, structure, variation, sequence\\
    & 39670 & cosmetics, color, skin, makeup, stain\\
    & 39122 & stimuli, patterns, response, signals, activation\\
    & 114454 & Beauty, identity, color, fashion, expression\\
    & 15678 & health, nutrition, immune, disease, metabolism\\
    \midrule
    \multirow{5}{*}{``what are the characteristics of the eucalyptus''}
    & 14689 & pets, veterinary, animals, dog, care\\
    & 15678 & health, nutrition, immune, disease, metabolism\\
    & 39122 & stimuli, patterns, response, signals, activation\\
    & 142071 & climate, weather, precipitation, seasons\\
    & 189083 & temperature, climate, humidity, weather\\
    \midrule
    \multirow{5}{*}{``best wr in nfl history''}
    & 69658 & wildcard, subsequences, activation, neuron\\
    & 71882 & baseball, athletes, performance, statistics\\
    & 78287 & classification, types, examples, varieties\\
    & 100445 & tennis, courts, justices, championships\\
    & 155393 & celebrity, entertainment, personality, humor\\
    \midrule
    \multirow{5}{*}{``how long is cough in children lasting''}
    & 15678 & health, nutrition, immune, disease, metabolism\\
    & 39122 & stimuli, patterns, response, signals, activation\\
    & 45139 & time, duration, sleep, hours, minutes\\
    & 56299 & measurements, values, dimensions, statistics\\
    & 185691 & weather, forecast, conditions, cold, outlook\\
\bottomrule
\end{tabular}
\end{table*}

\begin{table*}[t]
\small
\centering
\caption{\label{table:evaluation-minicpm} Reconstruction evaluation of sparse latent features and the reconstructed embeddings learned by our $k$-sparse autoencoder from \textsc{MiniCPM} embedding model.}
\begin{tabular}{l|cccc}
    \toprule
    &\multicolumn{4}{c}{\textsc{MsMarco}}\\
    &\textbf{MSE}&\textbf{MRR}&\textbf{P@10}&\textbf{R@10}\\
    \hline
    Original&--&0.3770&0.0682&0.6519\\
    \hline
    Sparse Latent (K=32)&--&0.1908&0.0389&0.3745\\
    Sparse Latent (K=64)&--&0.2594&0.0507&0.4870\\
    Sparse Latent (K=128)&--&\textbf{0.2953}&\textbf{0.0565}&\textbf{0.5416}\\
    Reconstructed (K=32)&0.00014&0.3128&0.0587&0.5613\\
    Reconstructed (K=64)&0.00011&0.3397&0.0630&0.6025\\
    Reconstructed (K=128)&\textbf{0.00009}&\textbf{0.3535}&\textbf{0.0649}&\textbf{0.6207}\\
    \bottomrule
\end{tabular}
\end{table*}

\begin{table*}[t]
\small
\centering
\caption{
\label{table:perspective:2-minicpm} 
Manipulation over for the binary perspective queries ``"What determines the success of rehabilitation therapy?'' and ``What shapes consumer decisions when buying eyewear?'' by amplifying the perspective latent features using \textsc{MiniCPM} as the embedding model.
}
\vspace{-0.2cm}
\begin{tabular}{c|p{32mm}|p{80mm}|c}
\toprule
    \textbf{Feature ID} &\textbf{Description from N2G} & \textbf{Retrieved Document} & \textbf{B/A} \\ 
    \midrule
    183 & energy, transformation, healing, vitality, balance & ``...Setting goals is the best way to achieve a successful rehabilitation outcome....''& 0/0\\
    4857 & time, duration, intervals, periods, estimation & ``With treatment, a few people recover in a year or less. For the vast majority, though, treatment and the recovery process take three to seven years, and in some cases even longer.'' & 0/5\\
    \midrule
    39423 & health, vision, care, eye, conditions  & ``What time of the day to have eye exam to get prescription eye glasses? I need a new pair of glasses (near sighted + other). I wonder it makes a little difference to go in the morning or afternoon or evening. I wonder if the eyesight is better in the morning after a night's sleep? Should I get eye exam when the eyesight is in best or worst condition?'' & 1/5\\
    161546 & glasses, eyewear, sunglasses, styles, features & ``When buying eyeglasses, the frame you choose is important to both your appearance and your comfort when wearing glasses. But the eyeglass lenses you choose influence four factors: appearance, comfort, vision and safety.''& 2/4\\ 
    \bottomrule
\end{tabular}
\end{table*}

\begin{table*}[t]
\small
\centering
\caption{
    \label{table:frequency-samples-minicpm}
    Sparse latent features from the frequency distribution using \textsc{MiniCPM} as the embedding model.
}
\vspace{-0.2cm}
\begin{tabular}{c|rl}
    \toprule
    \textbf{Region} & \textbf{Feature ID} & \textbf{Description from N2G}\\
    \midrule
    \multirow{6}{*}{Head}
    & 25 & health, medical, conditions, females, diagnosis\\
    & 97 & patterns, sequences, triggers, signals, behavior\\
    & 183 & energy, transformation, healing, vitality, balance\\
    & 197 & signals, patterns, thresholds, responses, stimuli\\
    & 207 & television, advertising, marketing, entertainment, engagement\\
    & 236 & ot, Rep, neuron, activation, subsequence\\
    \midrule
    \multirow{6}{*}{Torso}
    & 146050 & trading, hours, market, business, activities\\
    & 188194 & Health, recreation, arts, fitness, therapy\\
    & 140841 & health, wellness, community, education, environment\\
    & 109917 & health, wellness, nutrition, activities, rituals\\
    & 153312 & movie, technology, vehicle, animal, mechanics\\
    & 154625 & analysis, patterns, activation, signals, behavior\\ 
    \midrule
    \multirow{6}{*}{Tail}
    & 114226 & communication, education, resources, technology, collaboration\\
    & 107220 & health, wellness, genetics, lifestyle, information\\
    & 125167 & blood, language, difference, country, education\\
    & 144165 & cellular, biological, procedures, structures, metabolism\\
    & 193906 & neurobiology, stimuli, patterns, activation, response\\
    & 125701 & communication, processes, information, interactions, connections\\
    \bottomrule
\end{tabular}
\end{table*}

\begin{table*}[t]
\small
\centering
\caption{
\label{table:query-features-minicpm}
Top activated features from a subset of queries in \textsc{MsMarco} dataset using \textsc{MiniCPM} as the embedding model.
}
\vspace{-0.2cm}
\begin{tabular}{c|rl}
    \toprule
    \textbf{Query Text} & \textbf{Feature ID} & \textbf{Description from N2G }\\
    \midrule
    \multirow{5}{*}{``what is prism in eyeglasses''}
    & 161546 & glasses, eyewear, sunglasses, styles, features\\
    & 26168 & structure, geometry, prism, dimensions, properties\\
    & 39423 & health, vision, care, eye, conditions\\
    & 179744 & activation, patterns, sequences, neuron, inputs\\
    & 109256 & education, activities, science, culture, resources\\
    \midrule
    \multirow{5}{*}{``what are the characteristics of the eucalyptus''}
    & 47108 & neuron, activation, patterns, sequences, stimulation\\
    & 56389 & characteristics, organisms, life, description, taxonomic\\
    & 143997 & characteristics, features, descriptions, attributes, traits\\
    & 84508 & forest, trees, timber, ecology, sustainability\\
    & 134883 & Australia, Australians, territories, states, constitution\\
    \midrule
    \multirow{5}{*}{``best wr in nfl history''}
    & 16624 & football, NFL, teams, players, games\\
    & 179906 & receiver, wide, receptions, football, targets\\
    & 147634 & history, culture, documentation, information, analysis\\
    & 189070 & health, disease, communication, identity, experience\\
    & 143889 & patterns, sequences, neural, interactions, responses\\
    \midrule
    \multirow{5}{*}{``how long is cough in children lasting''}
    & 103545 & cough, symptoms, conditions, medical, causes\\
    & 29915 & children, pediatric, development, therapy, care\\
    & 174114 & lungs, breathing, pulmonary, respiratory, health\\
    & 4857 & time, duration, intervals, periods, estimation\\
    & 113082 & cough, chronic, symptoms, causes, prevalence\\
\bottomrule
\end{tabular}
\end{table*}

\begin{figure}[t]
\centering
\begin{subfigure}[b]{0.49\linewidth}
    \centering
    \includegraphics[width=\linewidth]{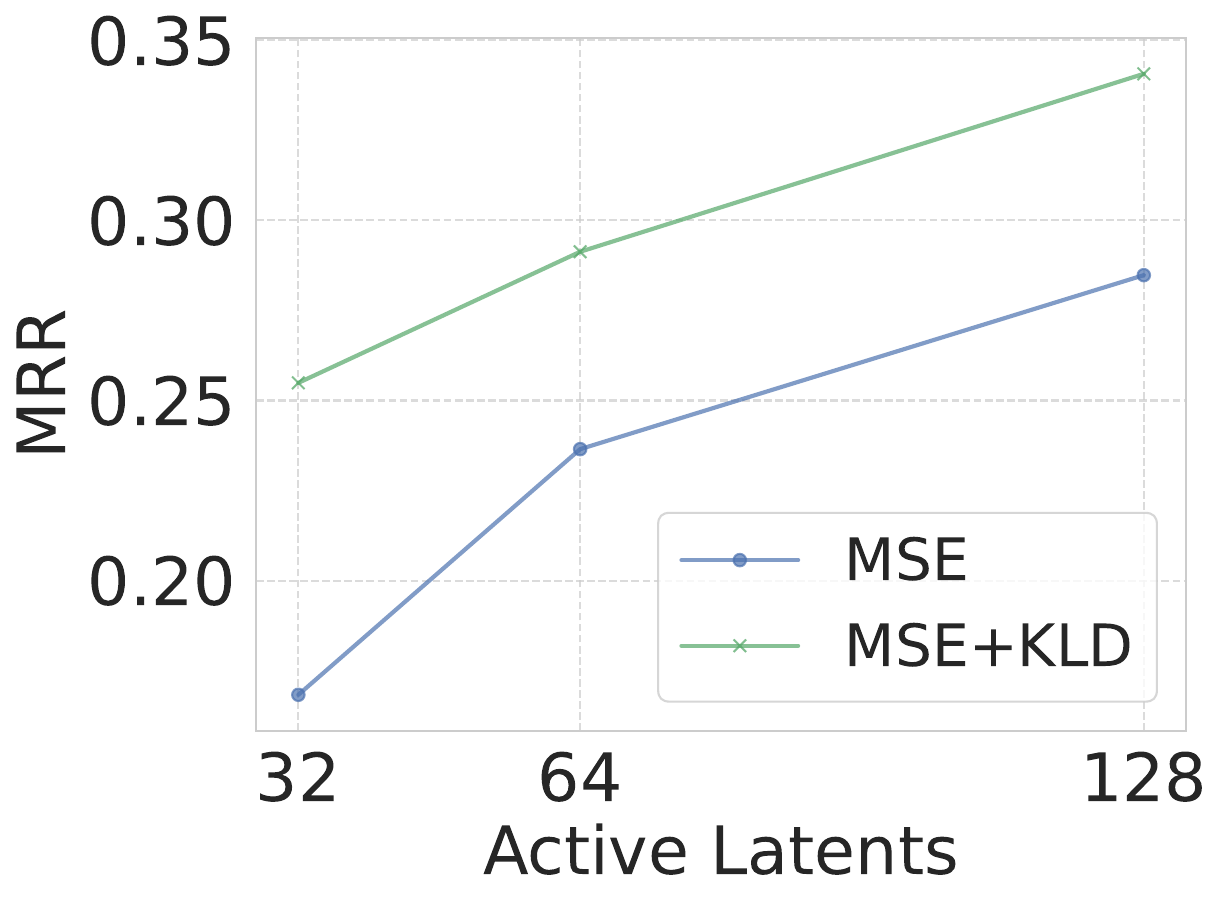}
    \caption{\textsc{BEIR} (Rec.)}
\end{subfigure}
\hfill
\begin{subfigure}[b]{0.49\linewidth}
    \centering
    \includegraphics[width=\linewidth]{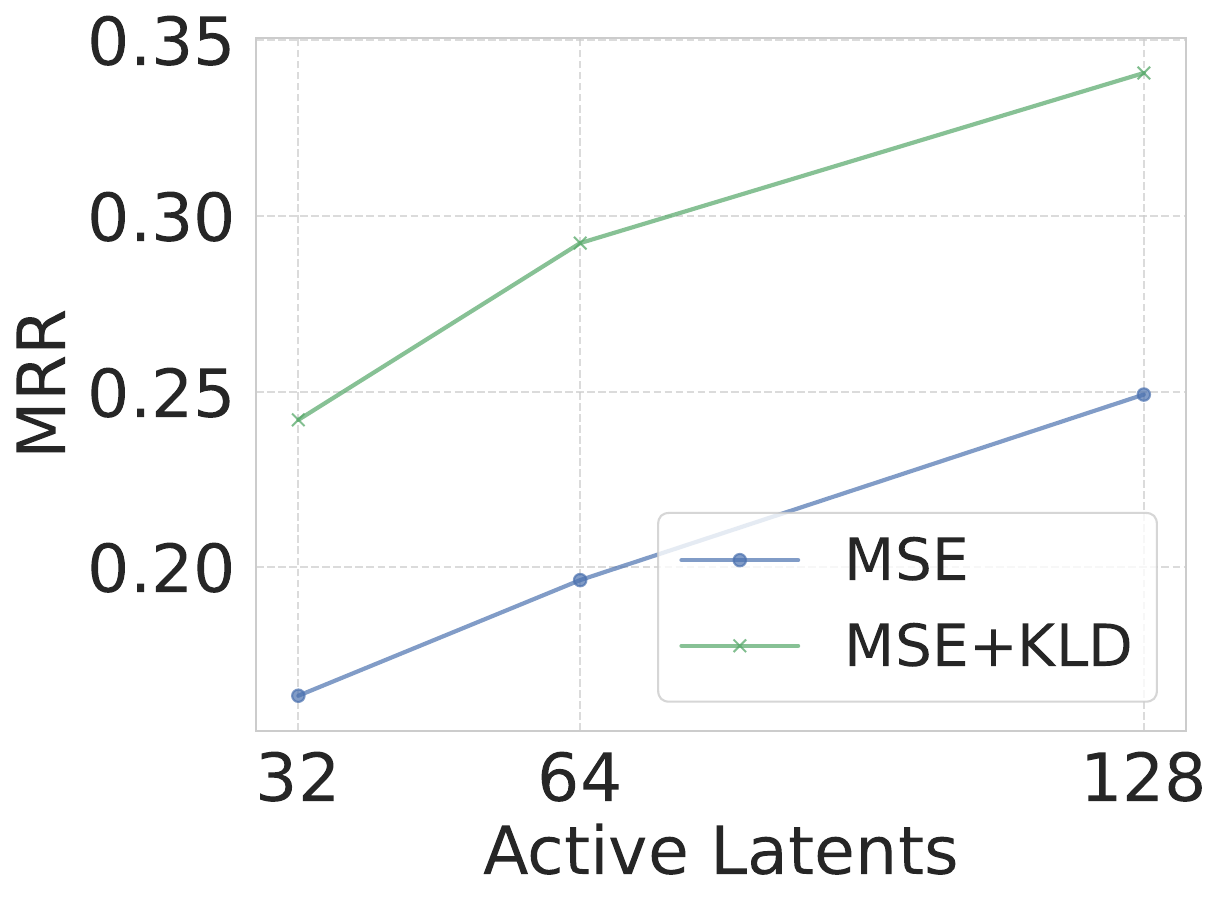}
    \caption{\textsc{BEIR} (Spr.)}
\end{subfigure}
\caption{\label{figure:ablation-beir} Retrieval performance of reconstructed (Rec.) embeddings and the sparse latent features (Spr.) before and after the contrastive loss KLD is applied on \textsc{Beir} using \textsc{Bge} as the embedding model.}
\end{figure}

\section{Training Procedures} \label{appendix:training}

During training, we employ the Adam optimizer \citep{kingma2014adam} with a batch size of 512 across 128 total epochs. The initial learning rate is set to $1 \times 10^{-3}$ and is progressively reduced using the cosine annealing scheduler \citep{loshchilov2016sgdr}. We sample 16 relevant documents per query from the original embedding space to compute the loss function in an efficient manner.

\section{Ablation Study} \label{appendix:ablation-study}

This section presents the ablation study, comparing models trained with MSE alone against those incorporating contrastive loss on the \textsc{Beir} dataset. The comparison is illustrated with Figure \ref{figure:ablation-beir}.

\section{Interpretability Study} \label{appendix:interpretability}

In our interpretability analysis, we utilize the N2G approach to interpret latent features extracted by the autoencoder. Sampled features from different parts of the frequency distribution (i.e. head, torso, tail) are shown in Table \ref{table:frequency-samples} along with their N2G explanations. Activated features and their associated semantic concepts for a subset of queries from \textsc{MsMarco} dataset are displayed in Table \ref{table:query-features}.

\section{Controllability Study} \label{appendix:controllability}

This section examines how feature activations can control retrieval on binary perspective queries. Tables \ref{table:perspective:1} presents how feature amplification affects the number of relevant documents retrieved before and after (B/A) over the binary perspective queries ``What is a key factor in the spread of infectious diseases?'' and ``What is a major influence on automotive emissions?''.

\section{Role of Base Embedding} \label{appendix:role-of-base-embedding}

This section explores the transferability of our method across different embedding models. As illustrated in Table \ref{table:evaluation-minicpm}, our approach demonstrates consistent performance when applied to the \textsc{MiniCPM} embedding. However, we observe a noticeable decline in retrieval accuracy when using the sparse autoencoder with $K=32$ active features. This reduction may be attributed to the significantly larger embedding dimension involved, which is three times the size of \textsc{BgeBase}. This increased dimensionality likely necessitates a greater number of active features to support the retrieval task. Additionally, the results of our interpretability analysis and controllability study, conducted using the \textsc{MiniCPM} embedding, are presented in Tables \ref{table:perspective:2-minicpm}, \ref{table:frequency-samples-minicpm}, and \ref{table:query-features-minicpm}.

\end{document}